\documentclass{article}
\usepackage{authblk}
\usepackage{geometry}
\usepackage{hyperref}
\usepackage{tikz}
\usepackage{graphicx}
\usetikzlibrary{shapes.geometric, arrows.meta, positioning, fit, backgrounds}
\usepackage{booktabs} 
\usepackage{caption}  
\usepackage{array}    
\usepackage[utf8]{inputenc}
\title{\textit{Sandpiper}: Orchestrated AI-Annotation for Educational Discourse at Scale}
\author[1,2]{Daryl Hedley}
\author[1,2]{Doug Pietrzak}
\author[1,2]{Jorge Dias}
\author[1,2]{Ian Burden}
\author[1,3]{Bakhtawar Ahtisham}
\author[1,3]{Zhuqian Zhou}
\author[1,3]{Kirk Vanacore}
\author[1,3]{Joshua Marland}
\author[1,3]{Rachel Slama}
\author[1,4]{Justin Reich}
\author[1,5]{Kenneth Koedinger}
\author[1,3]{René Kizilcec}

\affil[1]{National Tutoring Observatory}
\affil[2]{FreshCognate}
\affil[3]{Cornell University}
\affil[4]{Massachusetts Institute of Technology}
\affil[5]{Carnegie Mellon University}
\date{2026}

\begin{document}

\maketitle

\begin{abstract}
Digital learning environments now generate complex human-AI discourse and rich data on learning and instruction. However, traditional qualitative analysis remains a labor-intensive bottleneck that limits scale. We present \textit{Sandpiper}, a mixed-initiative system that connects large conversational datasets to human qualitative expertise. The platform pairs researcher dashboards with agentic large language model (LLM) engines to scale analysis without sacrificing rigor. \textit{Sandpiper} addresses barriers to AI adoption in education on secure university infrastructure. It also uses schema-constrained orchestration to reduce hallucinations and enforce adherence to qualitative codebooks. An integrated evaluation engine benchmarks AI performance against human labels and supports iterative model refinement and validation. This paper describes \textit{Sandpiper}'s core design goals and the features and functionality built to meet them.
\end{abstract}


\section{Introduction}
Digital learning environments that support human and AI discourse now generate unprecedented volumes of semi-structured data, from tutoring transcripts and classroom recordings to peer collaboration. These corpora could reveal how learning and instruction unfold, but using them at scale remains difficult. Traditionally, human experts have conducted qualitative analysis through close reading and line-by-line coding, a process that is costly, labor-intensive, and prone to fatigue and inter-rater drift. As a result, much educational data goes unanalyzed, which narrows the reach of educational research.

To address these challenges, we introduce \textit{Sandpiper}, an interactive mixed-initiative system for high-throughput, AI-assisted qualitative analysis of educational discourse. \textit{Sandpiper} addresses the limits of fully automated black-box approaches by treating AI as a partner rather than a replacement for human expertise. The platform lets researchers retain control through prompt management and orchestration, which enforce codebook schemas during inference. By combining a secure university-hosted large language model (LLM) cluster with a robust evaluation engine, \textit{Sandpiper} helps researchers use generative AI without sacrificing privacy or rigor.

This paper describes the system’s interface and architecture and shows how it provides a scalable, rigorous infrastructure for computational education research. Table \ref{tab:dg} links \textit{Sandpiper}'s core design goals to the components that implement them. DG1 protects privacy through secure, university-hosted infrastructure and data storage. DG2 improves reliability through schema-constrained orchestration that enforces qualitative codebook structure and validates annotation outputs during inference. DG3 supports verification through an evaluation engine that compares AI-generated labels with human annotations. Together, these goals define the platform’s central aim: scaling qualitative analysis without sacrificing privacy, rigor, or researcher oversight.

\begin{table}[ht]
\small 
\centering
\caption{Detailed Mapping of Design Goals, Descriptions, and Proposed Solutions}
\label{tab:dg} 
\begin{tabular}{>{\raggedright\arraybackslash}p{3.2cm}>{\raggedright\arraybackslash}p{4.2cm}>{\raggedright\arraybackslash}p{7cm}}
\toprule
Design Goal & Description & Solution \\
\midrule
\textbf{DG1: Privacy Protection and Data Security.} & Guarantee the privacy of educational data through scalable methods. & Secure, university-housed LLM cluster use and data storage. \\
\addlinespace
\textbf{DG2: Schema-Constrained Reliability.} & Prevent LLM hallucinations and malformed tags during the coding process. & LLM agentic orchestration to ensure rigorous enforcement of data structure and qualitative codebook schemas and annotation validation during inference. \\
\addlinespace
\textbf{DG3: Verification and Benchmarking.} & Support iterative refinement and validation of automated annotation. & Dedicated evaluation engine assesses alignment and classification accuracy of AI and human labels. \\
\bottomrule
\end{tabular}
\end{table}

\section{Related Work}
\label{sec:literature_review}

\subsection{Prior Systems for Qualitative Coding}

Our work draws on research in human-computer interaction (HCI) and computational social science. As qualitative datasets grow larger and more complex, HCI researchers have increasingly explored \textit{mixed-initiative systems} that extend human analysis without displacing human judgment \cite{horvitz1999principles}. Traditional qualitative data analysis (QDA) software, such as NVivo and MAXQDA \cite{kuckartz2014qda}, supports manual coding well but does not scale easily to thousands of transcripts. By contrast, fully automated natural language processing (NLP) approaches often miss the nuance of complex educational discourse and produce results that do not fit rigorous qualitative methods \cite{nelson2020computational}.

Large language models (LLMs) have changed how researchers approach qualitative coding. Tools such as Label Studio \cite{tkachenko2020label} and other specialized frameworks use LLMs to suggest codes and pre-annotate datasets \cite{Xiao2023,Karkera2023}. However, these systems still struggle to follow complex, researcher-defined codebooks. Hallucinations and malformed outputs add noise and force researchers to clean AI-generated data instead of analyzing it \cite{bubeck2023sparks}.

\subsection{Qualitative Coding Educational Discourse using LLMs}
Using LLMs in educational discourse analysis shifts AI from a simple automation tool to a "critical partner" in quantitative ethnography and computational social sciences. Recent studies show that LLMs can help researchers generate, refine, and consolidate complex codebooks for learning interactions \cite{Barany2024_CodebookDev}. However, performance often drops when systems move from codebook development to independent labeling, and that drop depends heavily on context. For example, GPT-4 can approach human reliability when coding virtual tutoring transcripts, but it still struggles with underrepresented or subtle constructs \cite{Liu2024_PotentialLimits, Liu2025_GPT4Better}.

In more complex settings such as whole-class discussions, LLMs often miss specific instructional moves and achieve only moderate agreement with expert coders \cite{Long2024_ClassroomDialogue}. These gaps point to a central challenge: reliable performance depends on strong "gold standard" data. Research suggests that errors in human-labeled baselines can distort model evaluation \cite{Nahum2025_LabelErrors}, and without high-quality labels, even expert prompting may not improve performance \cite{He2025_PromptingDark}.

Few-shot prompting can improve LLM performance in educational tasks, but it is not a cure-all \cite{Liu2025_GPT4Better,Geathers2025_OSCEBenchmark}. Misaligned examples can hurt performance when they do not fit the pedagogical context of the discourse \cite{Long2024_ClassroomDialogue}. These findings justify \textit{Sandpiper}'s design: a rigorous, schema-constrained infrastructure that treats prompt engineering and model evaluation as scientific work rather than black-box procedure.

\section{\textit{Sandpiper}'s Design Solutions}
\textit{Sandpiper} addresses these limitations through \textit{schema-enforced orchestration}, which iteratively verifies prompts, data structures, and annotation outputs throughout the workflow. This process reduces malformed outputs and keeps annotations aligned with the qualitative codebook. By integrating an evaluation dashboard into the annotation pipeline, \textit{Sandpiper} also turns prompt engineering into a rigorous research task and lets researchers evaluate AI performance against human baselines. In this way, the system supports scalable LLM-based analysis without giving up the methodological rigor required for educational research.

\section{System Architecture and Implementation}
\begin{figure*}[h]
\centering
\begin{tikzpicture}[
    node distance=1.5cm and 2.5cm,
    backend/.style={rectangle, draw=black, thick, fill=blue!10, text width=2.8cm, align=center, rounded corners, minimum height=1.2cm},
    db/.style={cylinder, draw=black, thick, fill=green!10, shape border rotate=90, aspect=0.25, text width=2.2cm, align=center, minimum height=1.5cm},
    ui/.style={rectangle, draw=black, thick, fill=yellow!10, text width=4.5cm, align=center, rounded corners, minimum height=1.2cm},
    llm/.style={rectangle, draw=black, thick, fill=orange!10, text width=3.2cm, align=center, rounded corners, minimum height=1.2cm},
    io/.style={rectangle, draw=black, thick, dashed, fill=gray!10, text width=2.5cm, align=center, rounded corners, minimum height=1.2cm},
    arrow/.style={-{Stealth[scale=1.2]}, thick},
    double_arrow/.style={{Stealth[scale=1.2]}-{Stealth[scale=1.2]}, thick},
    dashed_arrow/.style={-{Stealth[scale=1.2]}, dashed, thick}
]

\node[ui] (dashboard) {Researcher Dashboard \\ (Session Explorer \& Evaluation Engine)};
\node[ui, below=2cm of dashboard] (worker) {Schema-Constrained \\ LLM Orchestrator};

\node[io, left=2.5cm of dashboard] (input) {Raw Educational \\ Transcripts};
\node[backend, left=2.5cm of worker] (api) {Backend API \\ (Node.js/Express)};
\node[db, below=1cm of api] (mongo) {Data Store \\ (MongoDB)};

\node[io, right=2.5cm of dashboard] (output) {Annotated Data \\ \& Metric Reports};
\node[llm, right=2.5cm of worker] (gateway) {Secure AI Gateway \\ (Cornell AI Gateway)};

\begin{scope}[on background layer]
    \node[draw=red, dashed, thick, inner sep=20pt, rounded corners, fill=red!5, fit=(dashboard) (worker)] (engine) {};
    \node[above=0.3cm of engine.north, text=red!80!black, font=\bfseries] {Mixed-Initiative Client \& Orchestrator};

    \node[draw=blue!60!black, dashed, thick, inner sep=16pt, rounded corners, fill=blue!4, fit=(api) (mongo)] (cornellservers) {};
    \node[above=0.3cm of cornellservers.north, text=blue!60!black, font=\bfseries] {Cornell Secure Servers};
\end{scope}

\draw[arrow] (input.east) -- node[above, font=\footnotesize] {Upload} (dashboard.west);
\draw[arrow] (dashboard.south west) to[bend right=30] node[pos=0.35, left, xshift=-5pt, yshift=15pt, font=\footnotesize] {Ingest} (api.north east);
\draw[double_arrow] (api.south) -- node[left, font=\footnotesize] {Sync} (mongo.north);
\draw[arrow] (dashboard.south) -- node[right, font=\footnotesize] {Enqueue Task} (worker.north);
\draw[arrow] (worker.east) -- node[above, font=\footnotesize] {Prompts} (gateway.west);
\draw[arrow] (gateway.south) to[bend left=25] node[below, font=\footnotesize] {JSON Responses} (worker.south);

\draw[arrow, rounded corners] (worker.south) |- 
    node[pos=0.70, below, yshift=-6pt, xshift=8pt, font=\footnotesize] {Store Validated Labels}
    (mongo.south east);
\draw[arrow, rounded corners] (mongo.north east) -- ++(0.75cm,0) |- node[pos=0.25, right, font=\footnotesize] {Retrieve Labels} ([yshift=-0.2cm]dashboard.west);

\draw[arrow] (dashboard.east) -- node[above, font=\footnotesize] {Export} (output.west);

\end{tikzpicture}
\caption{The \textit{Sandpiper} Pipeline. A central mixed-initiative loop tightly integrates the researcher dashboard and the schema-constrained orchestrator. The backend API and datastore are hosted within Cornell secure servers, while external model access is mediated through Cornell's AI gateway.}
\label{fig:architecture}
\end{figure*}
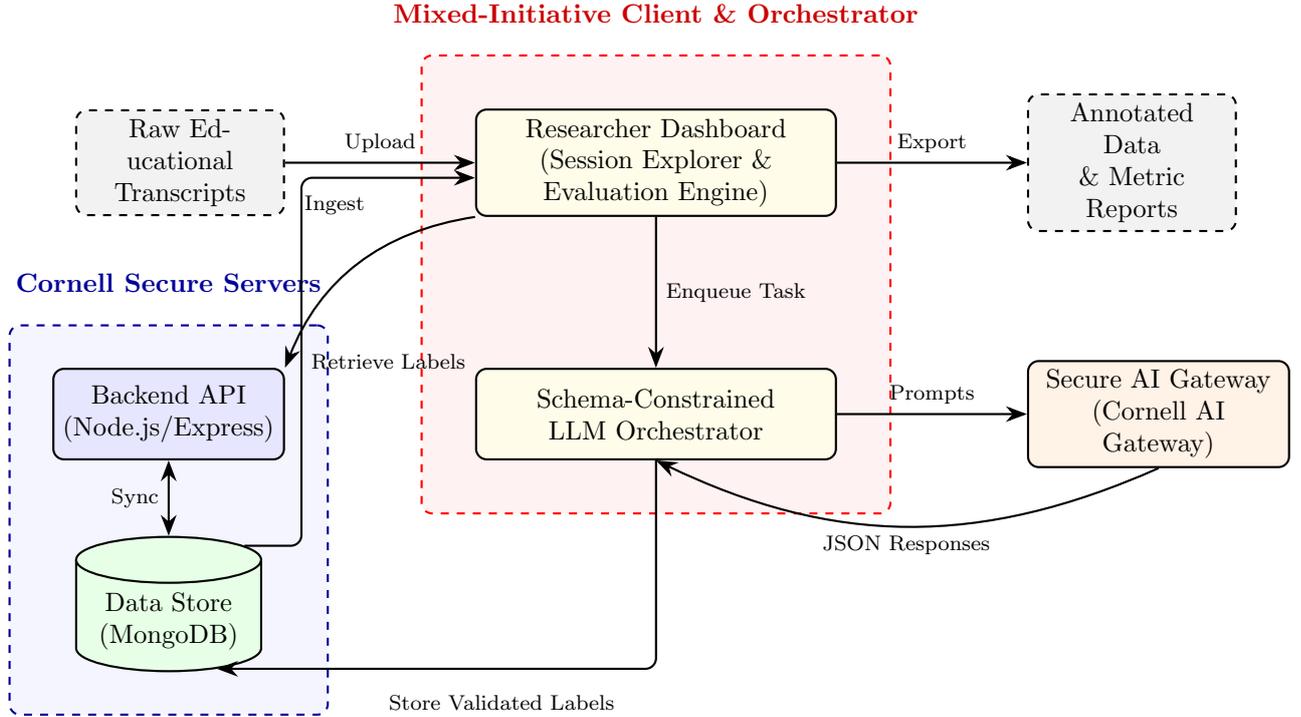

\textit{Sandpiper} uses a multi-tiered architecture in which researchers access orchestrated LLM workflows through an interactive dashboard, while the computational demands of inference remain isolated in the backend. Figure~\ref{fig:architecture} shows the system’s high-level data flow.

To understand how the platform realizes its Design Goals, we present a walkthrough of the core interface components and the human researcher's cognitive process.

\subsection{Privacy-Preserving Data Ingestion, Normalization, and LLM Use (DG1)}
\textit{Sandpiper} was designed to support LLM use that meets ethical guidelines and university Institutional Review Board (IRB) requirements. It stores data on Cornell's secure server infrastructure through Cornell Cloudification services\footnote{https://it.cornell.edu/cornell-cloud}
, and routes all LLM inference through Cornell's secure AI Gateway cluster. This architecture protects educational data while giving researchers access to closed models such as GPT, Gemini, and Claude without compromising institutional control over data.

\subsection{Prompt Management and Schema Enforcement (Addressing DG2)}
\textit{Sandpiper} treats prompts as part of the research workflow, so researchers can revise them while preserving documentation and version control. This record lets researchers see which prompt produced each annotation run and its results. Researchers can also mark prompts as “Production” once they meet a defined threshold, signaling that they are ready for use at scale.

\begin{figure}[htbp]
  \centering
  \includegraphics[width=.9\linewidth]{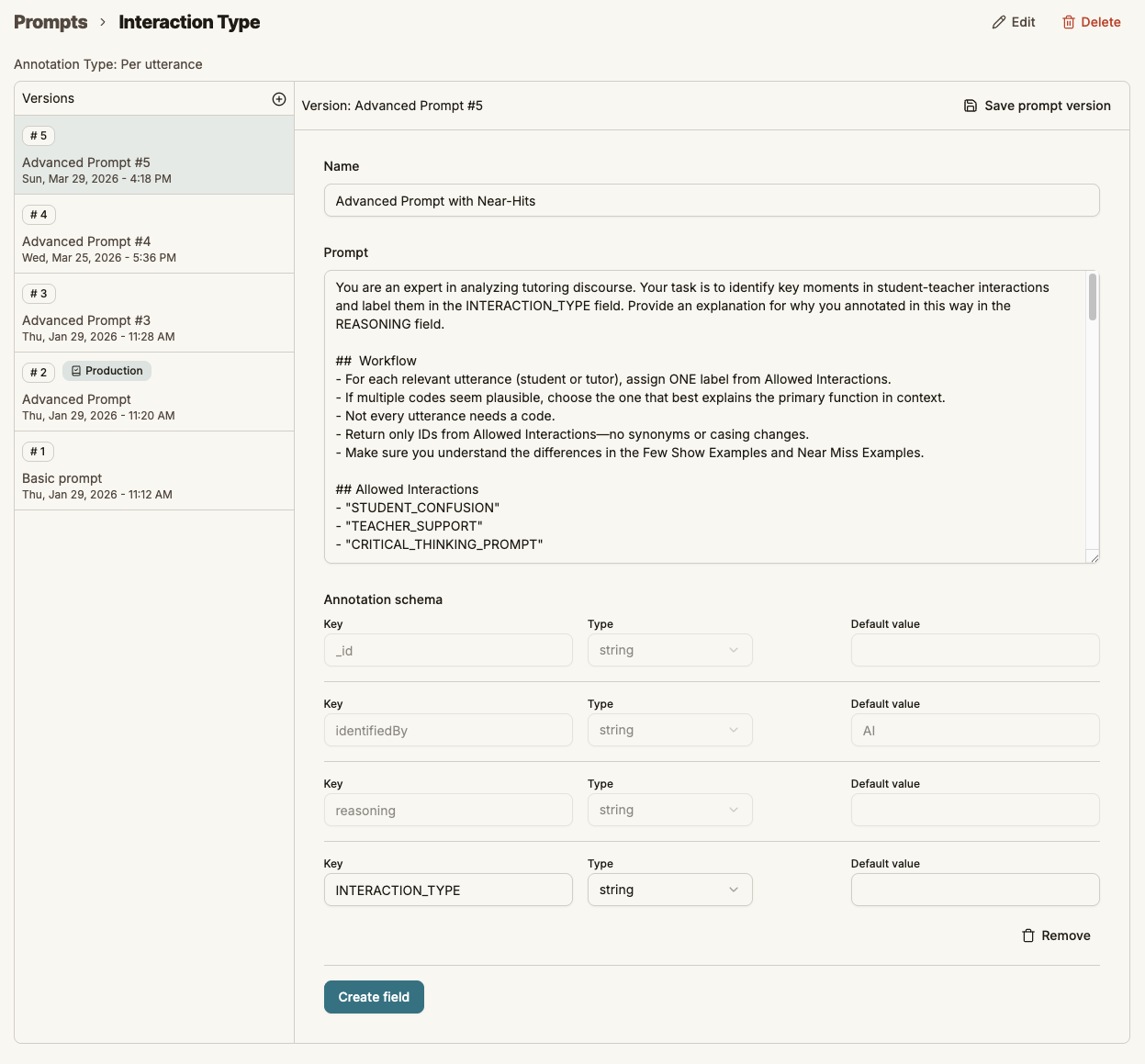}
  \caption{The Prompt Editor interface, illustrating version control for instructions and explicitly defined JSON coding schemas.}
  \label{fig:prompt_editor}
\end{figure}

\begin{figure}[htbp]
\centering
  \centering
  \includegraphics[width=.5\linewidth]{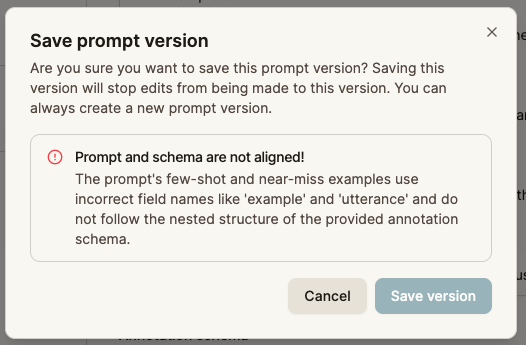}
  \caption{System mechanisms for ensuring inference output perfectly aligns with the target qualitative schema.}
  \label{fig:prompt_alignment}
\end{figure}

In the Prompt Editor, researchers define coding schemas alongside natural language instructions (Figure~\ref{fig:prompt_editor}). To address common LLM failures such as hallucinated or malformed tags, \textit{Sandpiper} uses a schema-constrained annotation engine. When background workers send LLM API requests, they use an \textit{Orchestrator Loop}. Rather than accept raw output blindly, each worker checks it against the user-defined JSON schema. If the output is malformed, the orchestrator generates feedback about the formatting error and re-prompts the LLM (Figure~\ref{fig:prompt_alignment}). This iterative process helps keep returned annotations aligned with the researcher’s qualitative codebook.

\subsection{Annotation Workflows and the Chat View}
Researchers steer the AI by defining runs: they select a corpus of sessions, a prompt version, and a target LLM. As asynchronous workers process the dataset, they attach structured JSON outputs directly to individual utterances.

\begin{figure}[htbp]
  \centering
  \includegraphics[width=.75\linewidth]{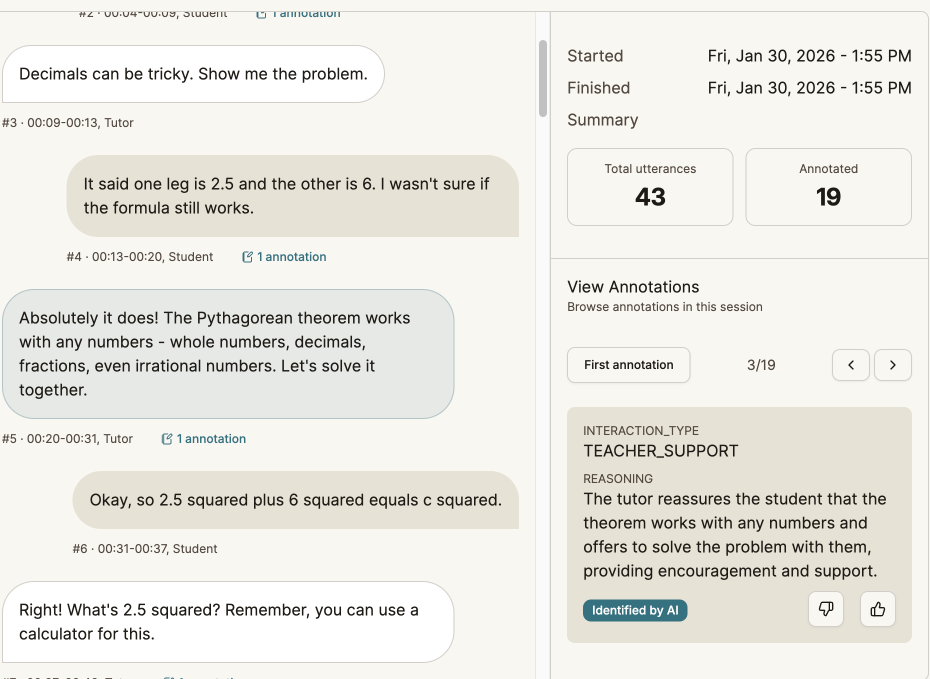}
  \caption{The Session Explorer interface detailing the Chat View and interconnected labeling panel.}
  \label{fig:chat_view}
\end{figure}

Within the platform’s Session Explorer, researchers use the Chat View and Labeling Panel (Figure~\ref{fig:chat_view}). This interface presents the original transcript alongside AI-generated codes. By viewing these labels next to the raw data, researchers can monitor the mixed-initiative process and identify cases where the codebook may need refinement.

\subsection{The Evaluations Dashboard (Addressing DG3)}
\begin{figure}[htbp]
  \centering
  \includegraphics[width=.9\linewidth]{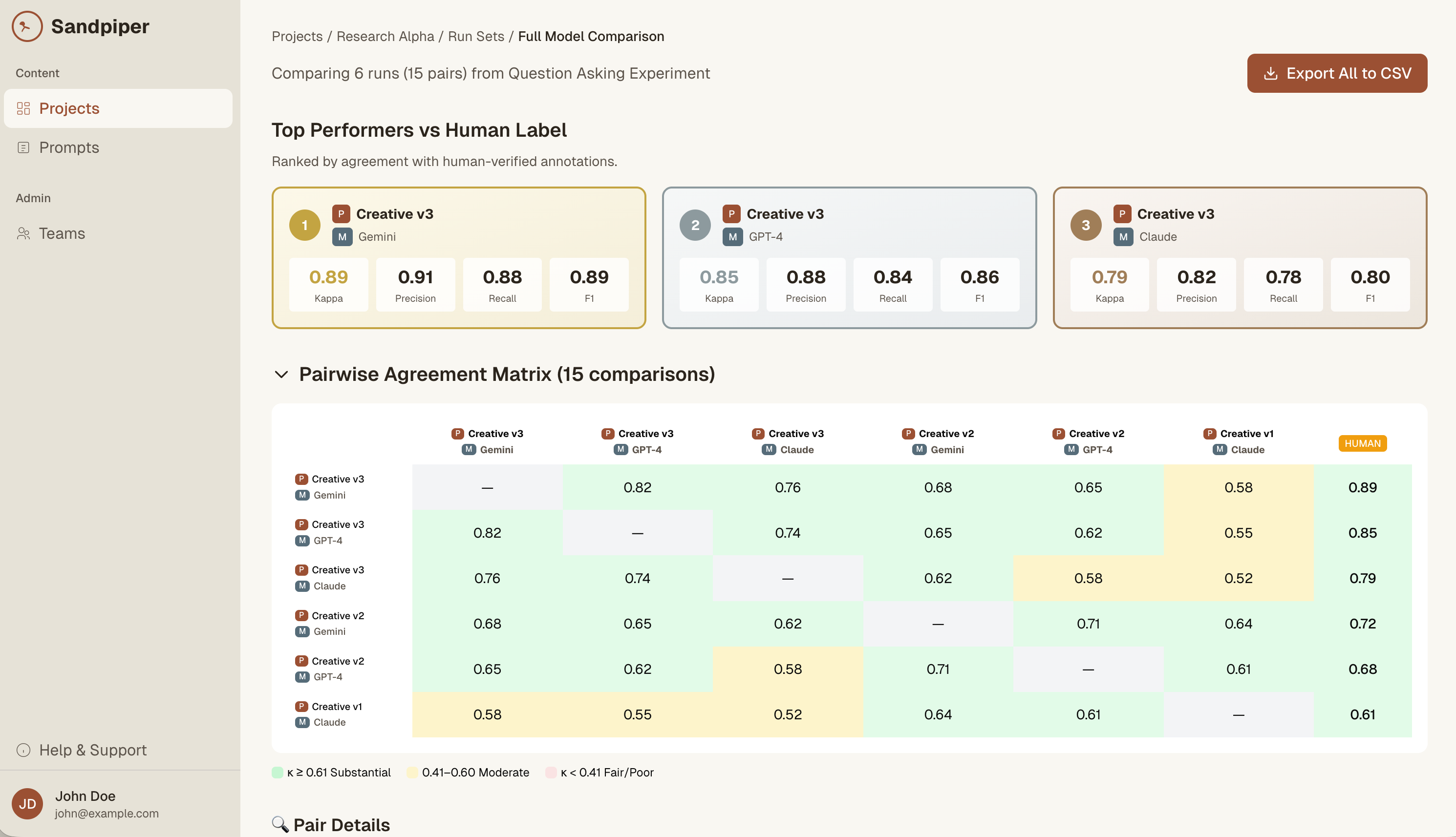}
  \caption{The Evaluations Dashboard aggregating metrics for multiple Experimental Run-Sets.}
  \label{fig:run_sets}
\end{figure}

\textit{Sandpiper}'s most distinctive feature is its integrated evaluation engine, which turns AI development into a rigorous qualitative research task. Researchers can group multiple runs into “Run-Sets” to compare annotations across models, prompt versions, and human labels. The Evaluations Dashboard (Figure~\ref{fig:run_sets}) computes key metrics, including pairwise agreement matrices, Cohen’s kappa, precision, and recall. By benchmarking schema-constrained LLM outputs against expert human labels, researchers can refine their codebooks and test the validity of the AI-assisted analysis.

\section{Conclusion and Future Work}
The \textit{Sandpiper} demonstrates a scalable, rigorous, and interactive approach to deploying LLMs for the qualitative analysis of educational discourse. By framing the system as a mixed-initiative platform equipped with schema-enforced inference and integrated evaluation tools, we empower researchers to scale their analyses without compromising methodological integrity.

Future work will expand the human-in-the-loop workflows, enabling researchers to seamlessly integrate manual corrections and dynamically chunk transcripts via ''Per-segmentation'' models prior to coding.

\bibliographystyle{plain}
\bibliography{references}

\end{document}